# A comprehensive analysis of the effect of ventilation and climatic conditions on covid-19 transmission through respiratory droplet transport via both airborne and fomite mode inside an elevator

Anish Pal[1], Riddhideep Biswas[1], Sourav Sarkar[1]*, Achintya Mukhopadhyay[1]
[1]Department of Mechanical Engineering, Jadavpur University, Kolkata, India
*Corresponding author: souravsarkar.mech@jadavpuruniversity.in

**Abstract**

Respiratory droplets that carry the disease-causing virus and are exhaled when speaking, coughing, or sneezing are one of the major causes of the worldwide Covid-19 epidemic. The risks of disease transmission due to these droplets are significantly high in confined public spaces like elevators. A numerical analysis using OpenFOAM has been performed in this work to investigate the droplet dispersion routes in an enclosed environment resembling an elevator. The effect of two scenarios on droplet dispersal, namely the quiescent and the fan-driven ventilation, both subjected to various climatic conditions (of temperature and humidity) ranging from cold-humid (15°C, 70% relative humidity) to hot-dry (30°C, 30% relative humidity and 30°C, 50% relative humidity) and hot-humid (30°C, 70% relative humidity) have been studied. A risk factor derived from a dose-response model based on the time-averaged pathogen quantity present around the passenger's mouth is used to quantify the risk of infection through airborne mode. It is found that the hot, dry quiescent scenario poses the greatest threat of infection (spatio-averaged risk factor 42%), whereas the cold humid condition poses the least risk of infection (spatio-averaged risk factor 30%). The humidity is discerned to emerge as 10 times more important factor in deciding the risk as compared to temperature. The proper Fan speed is determined for the epidemiologically safe operation of the elevator. The implementation of Fan ventilation at low RPM increases the risk as compared to a quiescent scenario, however with the increase in speed the risk decreases significantly. The Fan ventilation scenario with 1100 RPM (having a spatio-averaged risk factor of 10%) decreases the risk of infection by 67% in a hot, dry climatic condition as compared to a quiescent scenario. However, there is no significant reduction in risk beyond a certain Fan speed,1100 RPM. This Fan ventilation scenario (1100 RPM) is seen to sustain similar low risk irrespective of climatic variations. The deposition potential of aerosolized droplets in various parts of the respiratory tract namely the Extrathoracic and the Alveolar and Bronchial regions has been analyzed thoroughly because of the concomitant repercussions of infection in various depths of the respiratory region. Besides the airborne mode of infection, the fomite mode of infection (infection through touch) has also been investigated for both the ventilation scenarios.

**Keywords:** Respiratory droplets, Elevator, Risk Factor, Dose-Responsive Model, Climatic Conditions, Extrathoracic, Alveolar and Bronchial, Fomite, OpenFOAM,

## 1. Introduction

The continuing Covid-19 epidemic has resulted in a large number of deaths and has wrecked the global economy, with the whole world still struggling with its consequences. This pandemic was caused by the spread of SARS-Cov-2, and it has been established that the virus spreads through droplets generated through various respiratory activities.[1,2,3] Droplets expelled from the nose or mouth as individual talks, coughs, or sneezes have a various distribution of size and initial velocity. A thorough evaluation of the velocity distribution, ejection angle, and droplet size emitted from the mouth was investigated by Pendar et al[4] in order to determine the proper social distance criteria for various settings. Kwon et al.[5] used particle image velocimetry to determine the initial distribution of velocity in sneezing and coughing. The transmission and dispersion of these respiratory droplets largely depend on environmental conditions. Chong et al. implemented a direct numerical solution to calculate the lifespan of droplets under various ambient humidity levels[6]. They observed that at extremely high relative humidity (R.H.) (90 %), droplet lifespan might be prolonged up to 150 times. If a person is present in an open area like a school ground or marketplace, the risk of the disease transmission depends on wind velocity due to the available space between two persons. Feng et al. conducted a numerical investigation to evaluate the implications of relative humidity and wind on the transmission of respiratory droplets in an outdoor setting, and it was discovered that under the influence of wind, the microdroplets might travel longer distances[7]. Li et al. conducted a numerical evaluation on the dispersion of cough droplets containing non-volatile elements in a tropical outdoor setting, and the dispersion was impacted by humidity levels and wind velocity[8].

The aforementioned literature shows that ambient airflow characteristics affect the distribution of droplets in an open space. However, the issue gets much more critical in an enclosed place. In the first place, maintaining proper social distance may not always be feasible in small shared enclosed spaces like buses and elevators. In addition, in an enclosed environment, droplets emitted by a person would stay longer. Several investigations on droplet or aerosol dispersion in confined settings, such as classrooms[9], airline cabins[10,11], restaurants[12,13], buses[14], clinics[15], and courtroom[16] have recently been conducted. Cammarata et al. studied dynamic evaluation of the risk of viral infection in air[17]. Venkatram et al. used eddy diffusivity to model turbulent aerosol movement inside rooms[18]. Dbouk et al. demonstrated how changes in ventilation settings in enclosed places might affect airborne viral transmission[19]. Sen examined the transmitting of cough droplets in an enclosed domain such as a lift, taking into account a number of variables such as ventilation settings, the individual count within the elevator, the discharge direction, humidity level, and temperature[20]. Drivas et al. developed an analytical model that helps in assessing the human health risk from short-term indoor contaminants.[21] Xie et al. determined how far droplets can traverse in indoor environments[22]. Miller et al. investigated the superspreading event of Skagit Valley Chorale.[23] Ai et al. examined the spread of expiratory droplet nuclei in an indoor environment[24] Pal et al. studied the effect of climatic conditions on the dispersion of droplets in an indoor ambiance.[25] The droplet trajectories are determined for different ambient temperatures and relative humidities using a turbulent round jet model. Cheng et al. examined the paths of big respiratory droplets in different relative humidity in indoor environments[26]. The viral load dependency of the residence periods of virus-laden droplets from COVID-19-infected patients in indoor settings was studied by Srinivasan et al.[27] Biswas et al. reported the variation in dispersion of respiratory droplets with the variation of indoor environment and ventilation condition as well as the risk associated with the dispersion of droplets[28].

From the above discussion, we understand that an indoor domain poses a significantly higher risk of infection from virus-containing respiratory droplets as compared to an outdoor setting. Hence investigation of an indoor enclosed domain is very essential. In agreement with this developed understanding, in this current study, an elevator domain was chosen as the subject of investigation. There is a dearth of literature data on these aspects and hence has been the area of focus in this work.

- It is known that the virions are contained in the salt and protein-laden respiratory droplets. However, most numerical investigations have considered droplets to be made up of pure water, which does not portray a realistic scenario. Non-volatile salts in precise quantities are present in the droplets, and these salts encompass the virus. A salt-laden respiratory droplet will have thermophysical properties that are significantly different from that of a pure water droplet, and these thermophysical properties will change with the evaporation of the water from the droplet. This variation and continuous evolution in thermophysical characteristics will result in a change in mass transfer number and hence mass loss rate through evaporation, resulting in a change in overall droplet dispersal as compared to pure water. The salt-laden droplets will finally convert into aerosol after evaporation of pure water (solvent). Quite a few studies done recently have revealed that the aerosol mode of virus transmission is supported by sufficient evidence[19, 29-30]. Aerosols typically have a size lesser than 20 microns, allowing them to linger in the air for longer periods. Furthermore, based on current research, it may be argued that the aerosol contains the virions that cause disease transmission[28]. As a result, to resolve the issue discussed above and make simulations and their accompanying findings more authentic, the present study has considered the respiratory droplets to be made up of water and salt in certain proportions[31]. The actual composition of a respiratory droplet varies significantly with factors like age, gender, and state of health of the emitting person. So, it is difficult to ascribe a unique set of properties to the droplets ejected by an infected person. However, it has been shown that a salt solution can act as a reasonable surrogate for respiratory droplets[25]. The difference in the evaporation and dispersion characteristics of the droplet of a dilute saline solution is within the range of variation exhibited by respiratory droplets from different sources[25].

- Several previous studies investigating the transmission risk associated with respiratory droplets in an enclosed domain have not undertaken a detailed epidemiological investigation. Biswas et al. tried to quantify the risk of infection based on the droplet count that might be inhaled by a susceptible person[28]. Sen has also followed a similar approach for risk quantification[20]. In reality, the risk of infection depends on the number of virions inhaled by a person and does not depend on droplet count only. For this reason, in this current study, for all formulations pertaining to the risk of infection, the virion count supposedly inhaled by the susceptible person has been considered. Moreover, the majority of the studies that have taken into account the virion count have employed the classical Wells-Riley model for evaluation of the risk of infection associated with the virus-containing respiratory droplets[32–34]. However, there are some major drawbacks of the Wells-Riley model. Firstly, the Wells-Riley model is built upon a basic assumption that the indoor environment under consideration is a well-mixed room. Therefore, in situations (such as the situations in our present study) with ventilations that are either unable to render a well-mixed environment or have a directional nature, this assumption loses its validity[35]. Secondly, the Wells-Riley model requires the backward calculated quanta generation rate data[36] which is not available for any unfamiliar ventilation scenario. Thirdly, the Wells-Riley model, through the backward calculated quanta generation rate, incorporates many implicit errors (like geometry, ventilation, infectious source strength) and hence the model cannot be accurately used in any generic situation[36]. Additionally, the dose-response model considers a threshold dose, only after inhaling which the infection can be initiated, whereas, in the Wells-Riley model, there is no such concept of a threshold quanta or a threshold dose. Liu et al. compared four risk assessment models namely the Exposure Risk Index, Intake Fraction, Improved Wells-Riley model, and the Dose-Response model, and concluded that the dose-response is the most accurate one[37]. A more authentic approach for evaluating the risk of infection is to employ the dose-response model since it is free of these errors that are inherently present in the Wells-Riley model. The dose-response model does not require the assumption of a well-mixed room and bases the infection risk calculation on the straightforward pathogen count. Furthermore, dose-response models can consider many of the external influencing factors explicitly (such as air-flow pattern, spatial distribution, and dispersion of pathogens) and hence incur fewer implicit errors[36]. Hence, in our current study, the stochastic dose-response model[36] is used in place of the Wells-Riley model owing to the above-mentioned drawbacks of the Wells-Riley model and the corresponding triumph of the dose-response model over these drawbacks, as discussed earlier.

- When the droplets are inside the confined space, they evaporate, turning into aerosol and allowing them to stay suspended in the air for longer periods. Furthermore, small-sized droplets not only have the highest potential in getting deposited in the alveolar and bronchial region[38] but also, speed up the infection process as the escape time of the virions out of the droplet is directly proportional to the square of droplet size[39]. The ramifications of infection depend largely on what portion of the respiratory tract the pathogens are deposited. Infections in the alveolar and bronchial regions i.e., in the lower parts of the respiratory tracts or the lungs are more severe than that occurring in the Extrathoracic region[40]. Hence comprehensive scrutiny of these most precarious droplets, which have evaporated into aerosols, and their locations are also made, thus leading to recognition of local hotspots within the elevator domain, which must be attempted to be avoided by other fellow passengers. Apart from a generic understanding of the overall Infection Risk, these above analyses enable us to estimate the severity of the infection and the swiftness of infection initiation and infection spread. Although the virions lose their viability with the loss of water from the respiratory droplets, the half-life of the virions being in orders of magnitude much greater[41] than the exposure time of the elevator, the loss of viability has not been taken into account.

- Whenever a person sneezes or coughs in small confined places, such as an elevator, the aerosol will stay in the enclosure either suspended in the domain or get deposited on surfaces or escape out. Hence, these droplets present a risk of infection to another person traveling in the elevator both through direct and indirect transmission routes[36]. The majority of studies assessing the infection risk from the respiratory droplets have only considered the airborne route of infection. Very few studies have attempted to investigate the threat posed by the indirect mode of transmission posed by fomites. Biswas et al[28]. attempted to present the risk of possible infection through the fomite mode by reporting the droplet count that gets deposited on the surfaces of the elevator. However, they did not quantify the risk of infection through a proper formulation. In this present study, we have investigated the infection transmission through fomite mode as well and tried to quantify the risk associated with it based on the dose-responsive model.

- Droplet transmission routes in an enclosed domain affect the airborne risk of infection. The evaporation and heat transfer characteristics of the respiratory droplets affect the droplet trajectories significantly[28]. The ambient conditions play a major part in the evaporation and heat-transfer characteristics, and hence ultimately on infection risk through airborne routes for reasons as discussed earlier. Different ambient and climatic settings were explored, taking into account both a quiescent ambient and fan ventilation. Previous studies have tried to ascertain how the ventilation condition affects the risk of infection in an enclosed domain. Biswas et al[28]. reported that an exhaust fan maintains a very low risk of infection in an elevator. Sen studied the effect of axial inflow of air-jet on droplet transmission[20]. Fan ventilation system is fairly ubiquitous in typical elevators. Hence, a solution in the form of fan ventilation with proper fan speed has been prescribed that ensures a very low risk of infection in the domain. The obtained fan speed was found to be consistent in maintaining a low-risk situation in all climatic environments as compared to quiescent ambient.

In the present study, a 3-dimensional simulation has been used to investigate droplet dissemination across an elevator and its evaporation as well as the risk it poses. Inside the elevator, there is an infected passenger who is coughing and injecting droplets into the domain. The droplets are considered to be composed of a saline solution containing 1% NaCl (salt) and 99% $H_2O$ by wt. Computational Fluid Dynamics (CFD) approach using an Eulerian-Lagrangian framework, utilizing opensource software OpenFOAM has been employed to simulate the above situation. The Risk of Infection is quantified by a parameter called Risk Factor, whose formulation is based on the stochastic Dose-Responsive model. The spatial variation of the Risk Factor, as well as the average Risk Factor in the domain, has been enumerated for various climatic conditions and ventilation scenarios, seeking the most conducive condition which suppresses the Infection Risk significantly. Because of the differential infection severity due to pathogen deposition at various depths of the respiratory tract, the deposition potential of aerosolized droplets across various depths of the respiratory tract, particularly the Extrathoracic, Alveolar, and Bronchial regions, has been intensively investigated. Since, besides the airborne mode of infection, the fomite mode also presents a significant pathway for disease transmission, this mode of infection has also been investigated based on a rigorous formulation.

## 2. Problem Formulation

### 2.1. Geometry

The domain under study consists of a typical 1.2 x 1.2 x 2$m^3$ elevator, found in small enterprises or housing facilities, spacious enough to accommodate five individuals, which is shown in Fig. 1. The reason for the investigation of a smaller elevator is in accordance with our previous discussion that the risk of infection in small enclosed domains is higher. A 0.48 m diameter circular mounting is provided at the elevator's top. This top circular mounting is subject to various boundary conditions described in the section following this one and in Table 1. This top mounting is crucial because it permits the different ventilation scenarios inside the elevator to be simulated. As per the settings studied in this work, two boundary conditions: one representing a quiescent environment and the other representing a fan at the top entrance were provided. Customary ventilation ports are located on the bottom section of the elevator side walls as per EN-81 code[28]. The passenger's features are represented by rectangular boxes (for the reduction in computational cost, time, and complexity, also the region of focus is far from the passenger), the passenger height being 1.75 m and mouth being at a height of 1.56 m[19]. Figure 2 is an isometric representation of the passenger, while also showing the passenger's location relative to the domain. The passenger's mouth, which ejects and releases the cough droplets into the domain, has been designed as a rectangular slot with width and aspect ratio of 40 mm and 8 respectively[19].

### 2.2. Initial and Boundary Conditions

Different conditions are applied to the top circular mounting, according to various scenarios, as detailed in Table 1. The two ventilation ports below are modelled as pressure outlets. All the walls and the passenger's boundary are considered as walls with no-slip velocity boundary conditions. The door remains closed in all the scenarios and hence in all these scenarios, the standard wall boundary conditions are applied for the door. The cough is injected in four discrete stages with the coughing phenomenon beginning at 1s and ending at 4.12s. One cough is injected at a velocity of 8.5 m/s for a period of 0.12s normal to the surface of the mouth and releases 1008 droplets of mass 7.7mg[19]. The Rosin-Rammler distribution is the initial size distribution of cough droplets in each stage, following a scale factor and shape factor of 80 µm and 8 respectively[29]. The passenger, who is presumed to be an afflicted person, has a higher body temperature of 38.4°C. Continuous breathing (exhalation and inhalation) from the passenger's mouth has been considered. Various ambient conditions, as enumerated in Table 1 are implemented for various ventilation settings. The exhalation air expelled from the mouth is considered to be at body temperature, with a relative humidity of 100% [19]. The expelled cough droplets are also considered to be at body temperature, their initial composition being a combination of 0.01 mass fraction of salt (NaCl) and 0.99 mass fraction of water[31]. Broadly, two different ventilation scenarios namely, quiescent and fan have been examined to understand the influence of air ventilation on the transmission and evaporation of the cough droplets. Droplets expelled in the domain as a result of the passenger's coughing would travel the domain, their motion being dictated by the predominant velocity field. Depending on the ambient temperature and humidity, a certain number of droplets may form aerosols. A few droplets may escape, a few may get adhered to the different elevator surfaces, and the remainder will continue to float in the domain for long periods.

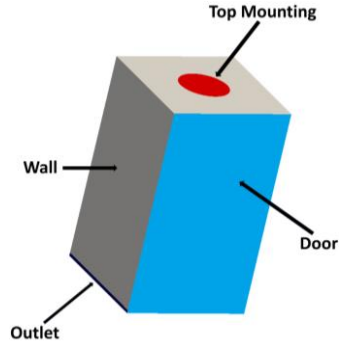 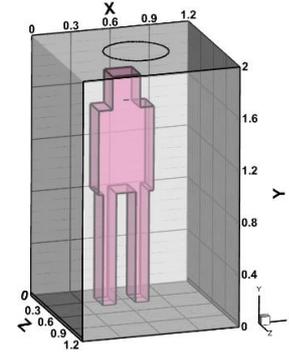

Figure 1: The entire computational domain.   Figure 2: Isometric view of the passenger and his location in the domain.

Table 1: Table of climatic Ambient environments along with Fan RPM, *ĵ indicates positive y-direction.

| Ventilation Scenario | Climatic Ambient environment | Angular Velocity of top circular mounting (rpm) |
|---|---|---|
| Quiescent | 15°C and 70% R.H. (cold humid) 30°C and 30% R.H. (hot dry) 30°C and 50% R.H. (hot dry) 30°C and 70% R.H. (hot humid) | 0 |
| Fan | 15°C and 70% R.H. (cold humid) 30°C and 30% R.H. (hot dry) 30°C and 50% R.H. (hot dry) 30°C and 70% R.H. (hot-humid) | -1100ĵ |

**Note:** the direction of fan rotation is in the anti-clockwise direction as seen (from below) by a person within the elevator. (Elevator Fans generally rotate in the anti-clockwise direction when seen from below)

## 2.3. Mathematical Model

The numerical model used in this study is the one recently developed by the authors[28]. For this numerical investigation, a fully coupled Eulerian-Lagrangian model was used. The Eulerian frame is used to model the carrier fluid, air. The continuity and the compressible multiphase mixture Reynolds-averaged Navier–Stokes equations in combination with the k – ω SST turbulence model(s) have been used for the carrier bulk multiphase fluid mixture[28,42]. The injected respiratory droplets are considered discrete particles and solved in the Lagrangian reference frame. The position of the droplets, as well as their velocity, are obtained considering the gravity, buoyancy, lift, and drag forces[28,42]. The Ranz-Marshall model[43,44] was used to compute the Nusselt number and the Sherwood number, to calculate the heat transfer and mass transfer of the droplets. The energy conservation equation provides the temperature of the droplet. Cough droplets are modeled as a mixture of water and NaCl (99 percent water and 1 percent NaCl by wt.)[31]. The mass fractions of the constituents alter when the droplets evaporate and lose their volatile liquid mass to the environment. Droplet attributes are governed by liquid water and salt properties, as well as their instantaneous mass fractions[28]. Droplet evaporation results in a continuous shift in the mass fractions of the components[28]. For this purpose, changes were implemented in the source code of OpenFOAM's reactingParcelFoam solver to implement this salt model of droplets. The detailed equations have been enumerated in the supplementary section of the paper for the kind perusal of any interested reader.

To solve the partial differential equations, the OpenFOAM (an open-source CFD-solver)[42] solver "reactingParcelFoam" was used, with the necessary adjustments to effectively apply the model for the salt solution as previously mentioned. It is essential to mention that each of the thermophysical features of both the Lagrangian and Eulerian phases is temperature-dependent. For its equation of state, the Eulerian phase has been characterized as an ideal gas, and its transport has been simulated using Sutherland's law for viscosity[45], which is based on the kinetic theory of gases and is applicable to non-reacting gases. Finite volume methods were used to discretize the Eulerian phase. Second-order schemes have been utilized for both space and time operators. Semi-implicit numerical techniques of second-order were employed for Lagrangian phase discretization.

In this current study, we are attempting to make our simulations more realistic by taking into account the influence of salt solution in cough droplets, i.e., by incorporating the salt model of droplets. Our salt model is in good agreement with the experimental results of Basu et al.[31] The overall numerical model, including the Eulerian model, the coupling between the Eulerian and Lagrangian models, the droplet heat transfer, and the evaporation model is validated against DNS data of Ng et al.[46] where a fair match between the two cases is noticed. These comparisons are reported and quantified elaborately in the previous work of the authors[28]. The plots pertaining to validation are also available in the supplementary material. A detailed grid-independent study has been performed, which to improve paper readability has been reported in the supplementary section.

## 3. RESULTS AND DISCUSSIONS

The dispersion and evaporation characteristics of respiratory droplets in an enclosed space were explored in this work. Both the phenomena have a significant impact on the epidemiological implications for the spread of virion-containing droplets. A detailed analysis of the epidemiological implications has been carried out using a dose-responsive model[36,47] which helps in developing a proper understanding of the infection risk associated with virus-containing cough-generated droplets. Firstly the droplet dispersion and its epidemiological implications in an elevator subjected to a hot dry climatic condition (30°C and 50% relative humidity, typically found in tropical and sub-tropical regions like Delhi in the early summer season) in a quiescent scenario have been investigated. Furthermore, the effect of fan ventilation on droplet dissemination as well as its epidemiological implications has been investigated, and the fan RPM that needs to be maintained for ensuring significantly low risk in the domain has been sought. Finally, the change in droplet evaporation and transport with the change in ambient conditions has been studied for both the fan ventilation and quiescent scenario. Besides involving the investigation of droplet transmission through airborne routes, the fomite route of infection has been investigated. All the studies have been considered for an elevator travel time of 10s, which is a typical travel time of an elevator for traversing 10 floors[28].

### 3.1. Transport, Evaporation, and Epidemiological implications of droplets in a quiescent scenario (30°C and 50% relative humidity)

The top mounting is considered as a wall in this scenario and with no enforced airflow within the domain, thus resulting in a quiescent condition within the elevator. The risk of infection from the ejected cough droplets within the elevator domain has been investigated. The domain has been sub-divided into 16 breathing boxes (each of size 0.3m x 0.4m x 0.3m) as illustrated in Fig. 3. The Risk of Infection(R) is calculated in all of these boxes using a dose-response model[36,47] as per equation 1 below.

$$R = 1 - \exp(-\sigma\mu) \quad (1)$$

μ is the expected number of pathogens likely to be inhaled by a susceptible person over the exposure time of elevator travel (10s), given by equation 2.

$$\mu = \sum_\beta \int_0^1 \frac{P}{V} \frac{\int_0^{T_0} \iiint_V \frac{c\pi}{6} D_0^3 N \, dt}{T_0} dt \quad (2)$$

where $D_0$ is the initial diameter of injected cough droplet; N is the mean viral load in the respiratory fluid of covid-19 infected persons (7 X 10$^6$ RNA copies/ ml)[48]. P is the pulmonary ventilation rate[49], V is the volume of the considered breathing box and β is the total number of inhalation cycles over the total exposure time ($T_0$), considering an inhalation period of 1s[28]. $\frac{P}{V}$ represents the probability of inhaling a droplet suspended within the breathing box. The infectivity factor, "σ", is the inverse of the number of viruses capable of initiating infection and is a direct indication of the infectious dosage[40]. Infectivity factor for SARS-COV-2 as reported by a recent study by Mikszewski et al[50] has been used for the current work. The dose-response model as discussed earlier is devoid of some of the limitations of the Wells-Riley model due to its consideration of the external factors like ventilation, geometry, air-flow pattern as well as the actual pathogen content.

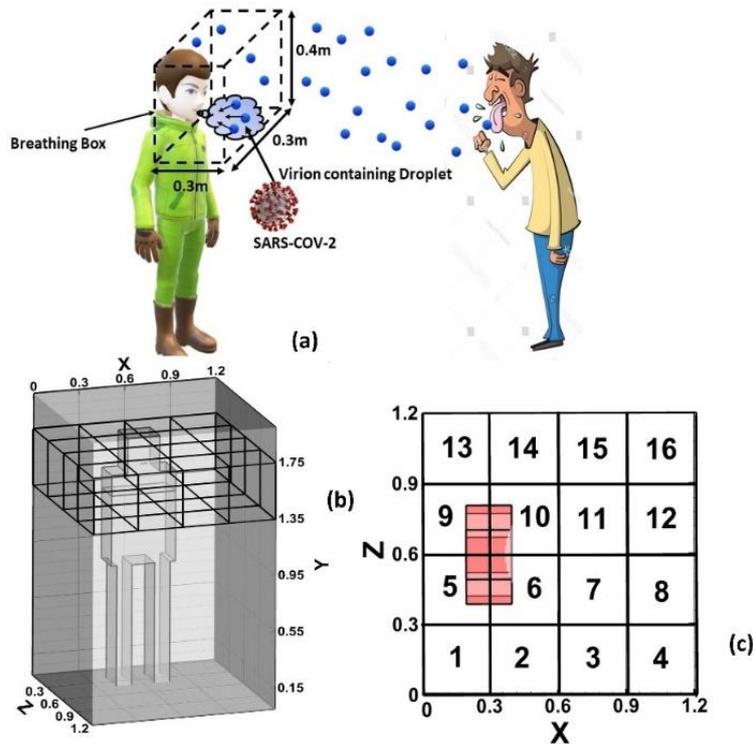

**Figure 3: (a) Illustrative representation of a breathing box containing virion-containing droplets and a susceptible person inhaling it.**
    **(b) An isometric view of all the sixteen breathing boxes in the domain at a height of 1.35m**
    **(c) Top view representing the passenger and location of all the sixteen breathing boxes with respect to the passenger.**

The droplets, in a quiescent domain, do not reach the floor during the specified 10-second elevator travel period as can be understood from Fig.4. It is crucial to notice that droplets do not fall immediately to the floor, but rather become entrapped in the turbulence caused by both the cough and the passenger's repeated intake and expiration, and spread throughout the elevator. The absence of any continuous draft of air in the domain slows down the process of sticking or escaping the droplets, hence the majority of them remain suspended in the air for a substantial amount of time. The probability distribution of the diameter of the suspended droplets at various time intervals, in Fig.5, shows that the suspended droplets evaporate continuously, causing them to shrink in size. The probability distribution is based on the total droplet count encompassing all four coughs. The likelihood of the formation of droplet nuclei is also depicted in Fig. 5. Droplet nuclei are droplets in which the liquid component has entirely evaporated and only the non-volatile components remain. We have droplet nuclei of different sizes (Fig. 5)) since their corresponding initial droplet size distribution follows a Rosin-Rammler distribution (as found in Fig. 5). A larger droplet may evaporate and lose its fluid content to become a nucleus of the same size as a droplet that is originally injected as a smaller droplet. It is worth mentioning in this context that, when considering a droplet and droplet nuclei of the same diameter, the droplet nuclei are more virulent. This is due to the fact that virions from an inhaled droplet nucleus can diffuse out into the host cell within a time span that is substantially less than that of a droplet with fluid content and having the same diameter[51], thus speeding up the infection process. After 10s, a considerable fraction of the suspended droplets evaporates, resulting in droplet nuclei ranging in size from 5 to 20 µm, as can be seen in Fig. 5. This phenomenon of suspended droplets evaporating to a size range of 5-20 µm produces a precarious situation in the elevator as this diameter range causes the droplets to remain suspended in the domain for extended periods[52]. Also, droplets having a diameter of less than 20µm have the highest penetration power in the lower part of the respiratory tract namely, the alveolar and bronchial region[38], thus causing more severe medical conditions

Figure 6 shows the magnitude of risk of infection and its spatial variation over all the 16 breathing boxes of the domain. It depicts that in a quiescent scenario over the exposure time of elevator travel, a substantial portion of the elevator domain presents a significant risk of infection. The calculated spatial-averaged risk factor (Spatial-averaged risk factor is the risk of infection(R), spatially-averaged over all the 16 breathing boxes) of 35.68% demonstrates that the quiescent scenario creates a high-risk condition inside the elevator. Furthermore, the hot and dry ambient environment leads to significant evaporation of the droplets leading to the development of a significant number of droplets in the size range of 5-20 µm. The droplets in this size range, virusols (droplet size<20µm)[53], derived from the relatively larger droplets have very high viral loads, less diffusion time of virions, and as discussed earlier due to their size being less than 20 µm have the highest potential for infection.

**3.2. Transport of droplets in a domain with a Fan mounted on the Top (30°C and 50% relative humidity)**

We infer from the preceding discussion that the absence of forced convection inside the elevator is unfavorable for enclosed domains because droplets remain suspended in the domain for extended periods, resulting in a very high-risk factor. As a result, we attempted to investigate the influence of forced convection on droplet dispersion characteristics by modelling the elevator's top mounting as a fan. The droplet dispersion in the domain is depicted for various fan speeds and their corresponding volume flow-rates[54,55] in Fig. 7. The risk of infection within the domain over the total exposure time was investigated for various rotational speeds of Fan. It has been found that at low values of Fan RPM, the risk of infection in the domain is more as compared to the quiescent scenario, however, with the increase in Fan RPM the risk of infection in the domain decreases, and at a particular RPM, it eventually becomes lower than the quiescent scenario, thus providing a pathway to a probable solution of risk mitigation. The spatial variation of Risk over the domain for various Fan speeds as well as the spatio-averaged risk of infection over the exposure time, depicted in Figs. 8 and 9 respectively reconfirm the same.

One interesting aspect pertaining to the droplet dispersion that needs to be noted is that the droplets upon injection from the mouth initially go up and then descend. The droplets being injected from the mouth get trapped in the vortices near the mouth and hence go up. The velocity vector plots (on plane AB, Fig. 10 (a)) in Fig. 10 (b)-(c) depict the vortices developed near the coughing passenger's mouth. As seen from Fig. 10(b), there is a weak velocity field and weak vortex created at low RPM (500 RPM), whereas there is a strong velocity field and strong vortex formed due to the enhanced induced turbulence brought about by the increasing rotating component of the fan at high fan speed (1100 RPM) Fig.10 (c). In the quiescent case, the droplets descend down slowly (but steadily) and orderly. The weaker vortices of the low-speed fan keep the droplets levitated for a substantial amount of time whereas the strong vortices of the high-speed fan carry the droplets away from the passengers and cause a significant percentage of droplets to stick on the domain walls. The droplet dispersion and its temporal history for various fan speeds depicted in Fig. 7 corroborate this argument. The above argument explains the increase in risk as we go from quiescent to low RPM Fan and the decrease in risk as we go from quiescent to high RPM Fan.

Thus we conclude that although at lower fan speeds, the risk is higher, increasing the fan speed ultimately decreases the risk of infection as compared to the quiescent scenario and the risk becomes significantly low, at a fan speed of 1100 RPM. However, increasing the speed beyond 1100 RPM yields only a negligible decrease in the risk factor. Hence, we conclude that maintaining a Fan speed of 1100 RPM results in very low risk (because further increasing the Fan speed does not yield major reductions in Risk Factor) and should be maintained in the domain to ensure maximum safety. Critically speaking, the hot dry (30°C and 50% relative humidity) environment and fan RPM of 1100 (spatio-averaged risk of infection–12%) brings about a 67% reduction in the risk of infection from that of a hot dry (30°C and 50% relative humidity) quiescent ambient environment (spatio-averaged risk of infection – 36%).

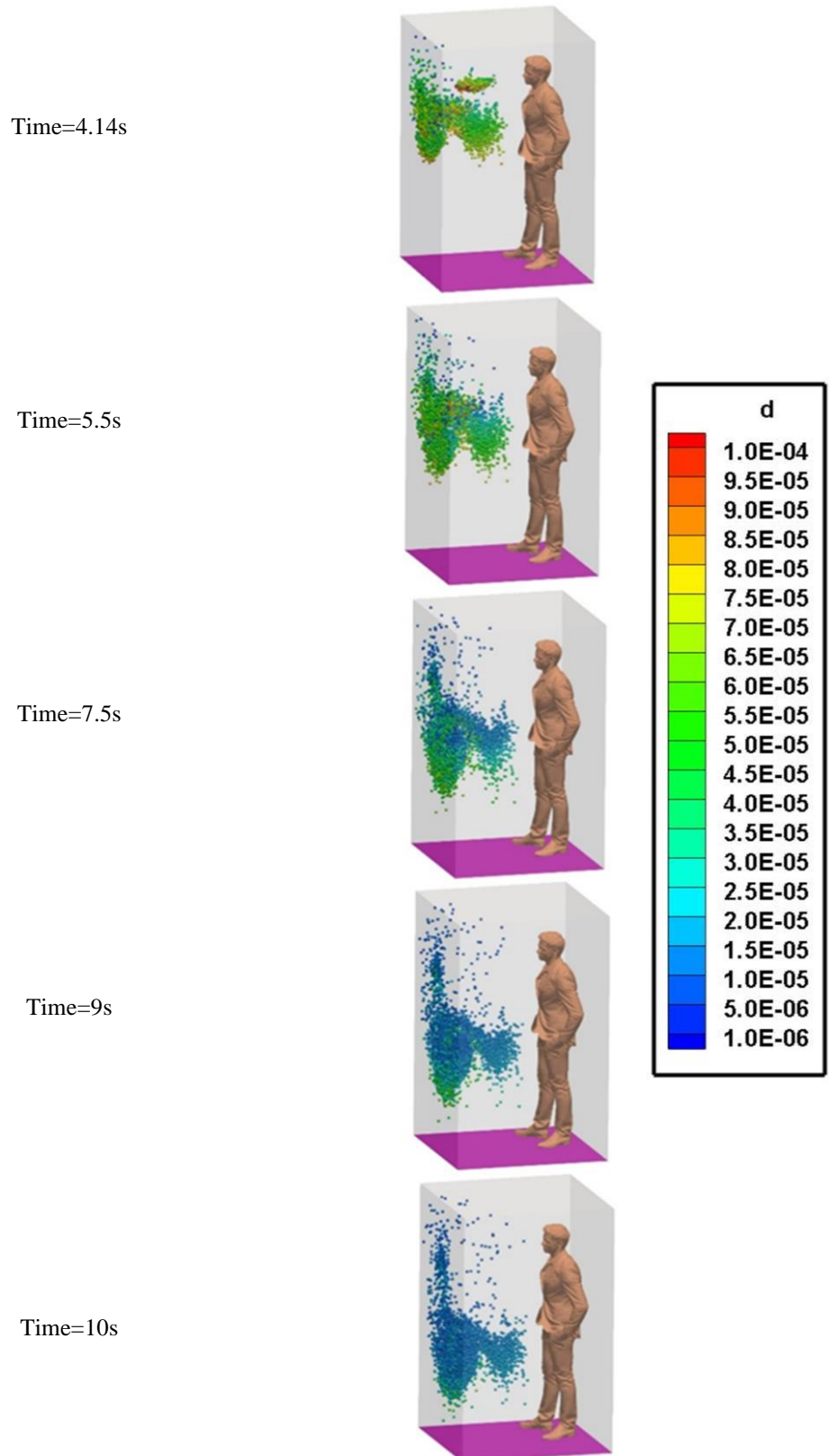

**Figure 4: Droplet dispersions at various time instants in a quiescent ventilation scenario of hot dry ambient environment (30°C and 30% relative humidity).
(Mannequin has been shown for illustration purpose only)**

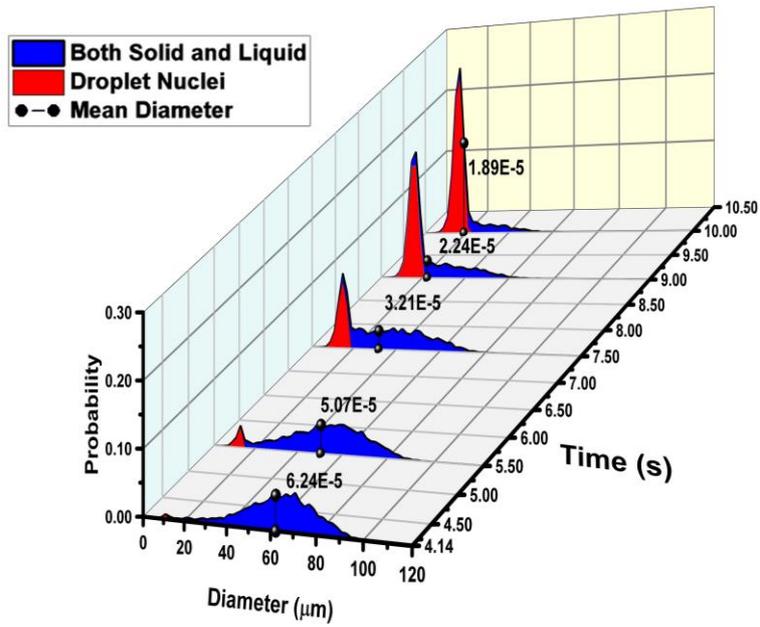

**Figure 5: Diameter distributions of suspended droplets at various time instants (4.14s, 5.5s, 7.5s, 9s, 10s.)**

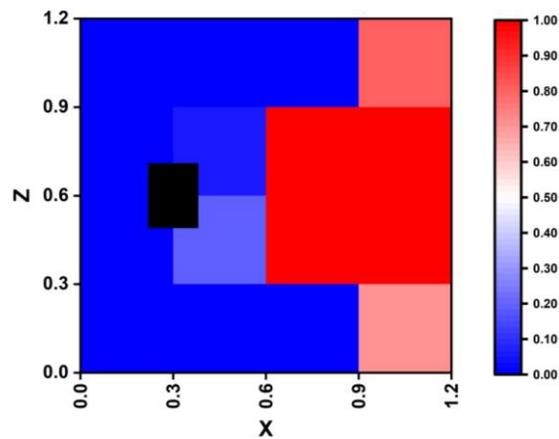

**Figure 6: Risk map showing the variation of risk of infection across the breathing boxes throughout the domain over the total exposure time for a hot dry quiescent ventilation scenario (30⁰C, 30% R.H.) (location of the infected passenger has been indicated by a black rectangular box).**

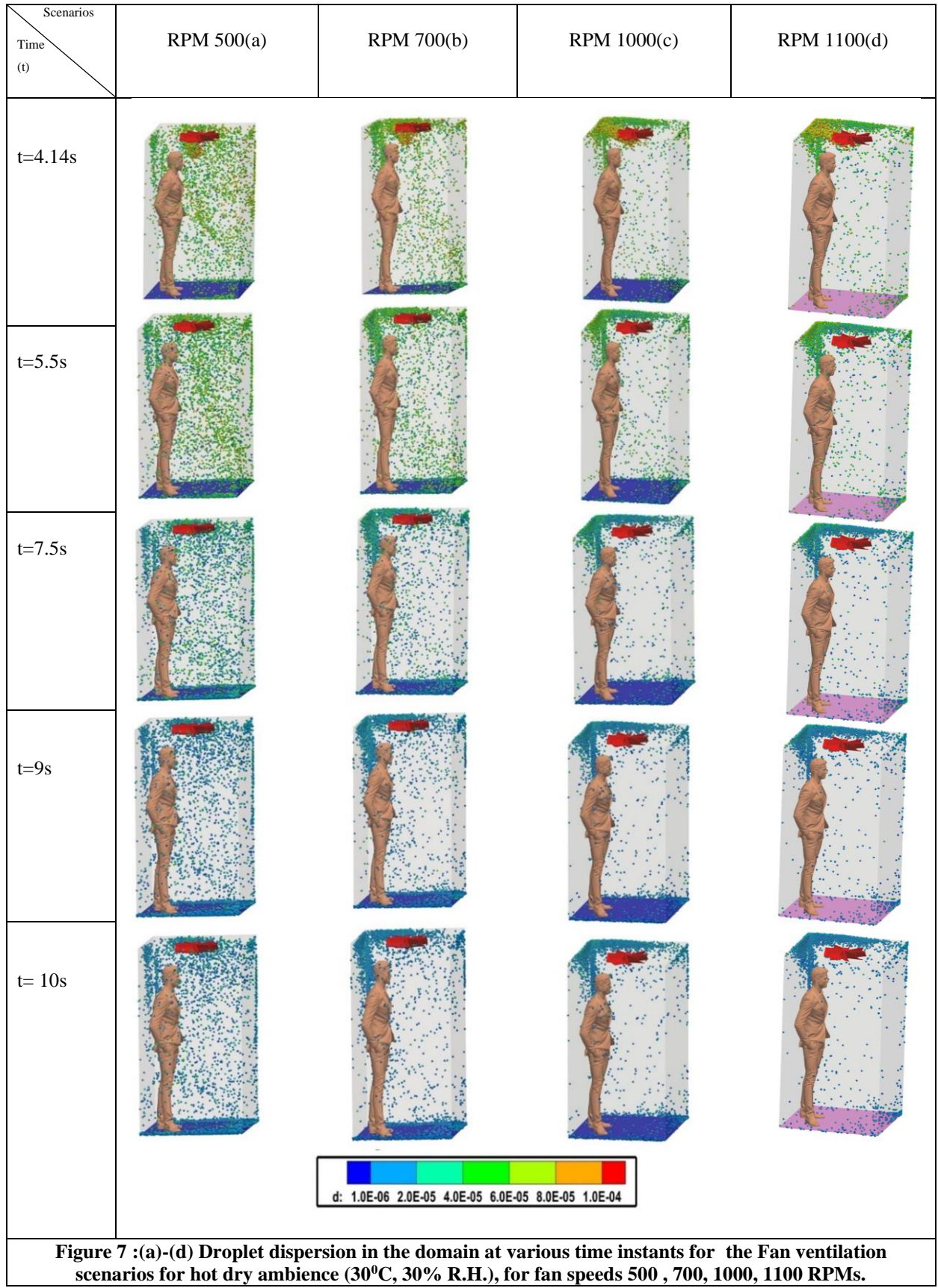

**Figure 7 :(a)-(d) Droplet dispersion in the domain at various time instants for the Fan ventilation scenarios for hot dry ambience (30$^0$C, 30% R.H.), for fan speeds 500 , 700, 1000, 1100 RPMs.**

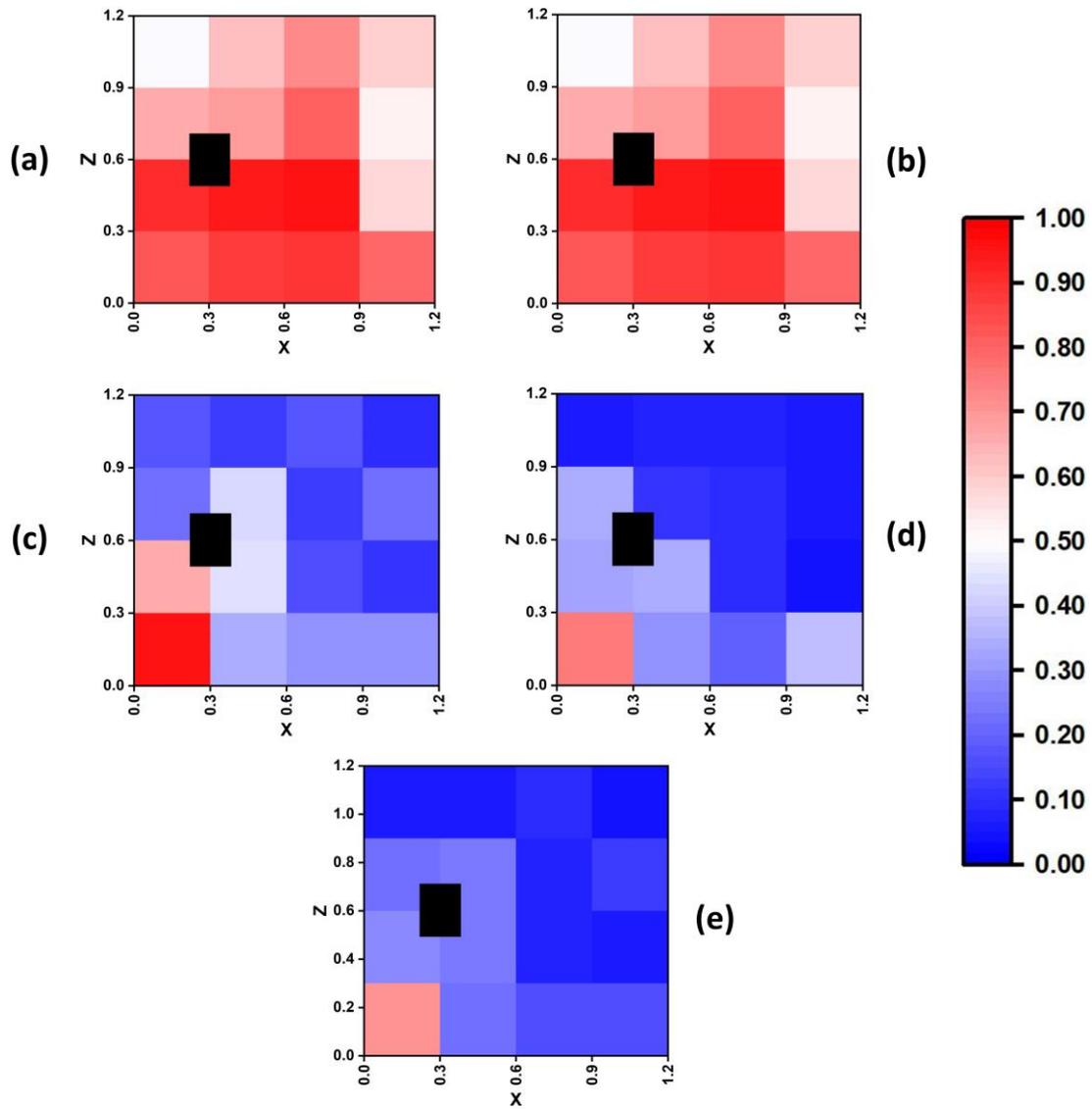

**Figure 8:** (a)-(e)Heat map showing the variation of risk of infection across the breathing boxes throughout the domain over the total exposure time for fan rpms 500,700,1000,1100,1200 respectively in a hot dry (30°C, 30% R.H.) ambient. (location of the infected passenger has been indicated by a black rectangular box).

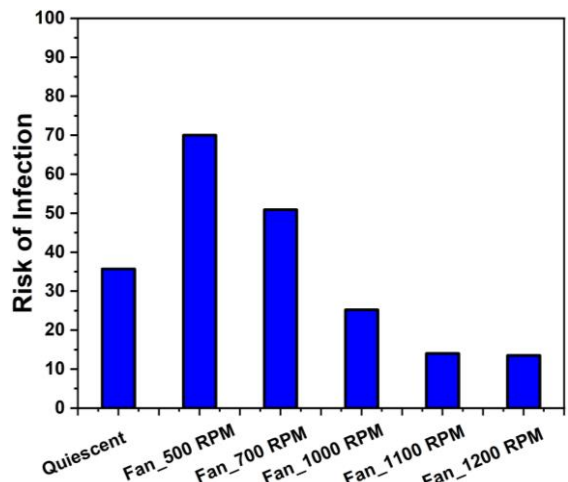

**Figure 9:** (a)-(e) Comparison of spatio-averaged risk of infection between quiescent and various fan rpms in a hot dry (30°C, 30% R.H.) ambient.

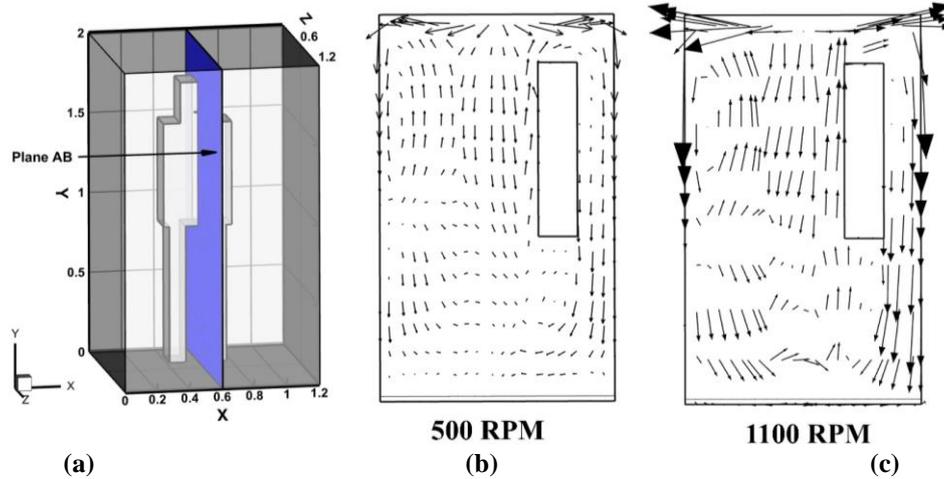

**Figure 10 (a)-(c): Plots the velocity vectors for various Fan velocities on Plane AB(Fig. 10(a)).**

**3.3. Effect of Climatic conditions on droplet dispersion and epidemiological implications in both Quiescent and Fan Scenarios.**

We were able to ascertain that the quiescent scenario offers a major risk in a hot, dry environment. This is due to the droplets evaporating rapidly, causing them to lose substantial mass and remain suspended in the domain, hence raising the risk of infection. But to develop a comprehensive understanding of this ventilation scenario, we must explore the influence of climatic variables on droplet transmission routes in an enclosed domain under quiescent condition. The effect of a wide range of climatic conditions varying from cold humid (15°C, 70% R.H.) to hot dry (30°C, 50% R.H and 30°C, 30% R.H) and hot humid (30°C, 70% R.H) conditions on the droplet dispersion characteristics along with their epidemiological implications have been investigated. Furthermore, the previously determined fan speed of 1100 RPM for the hot dry climatic condition for minimizing risk has been implemented in various climatic circumstances to understand if it can sustain low risk in all of the ambient conditions listed in Table1. After investigating the above wide range of climatic conditions, two extreme climatic conditions were identified – the cold humid climatic condition with 15°C temperature and 70% R.H. and the hot dry climatic condition with 30°C temperature and 30% R.H. The Infection Risk, evaporation characteristics and other related parameters for the rest of the climatic conditions were found to lie in between the two extreme climatic conditions. Hence, the main focus has been concentrated on these two extreme climatic conditions.

**3.3.1. Transport and Evaporation of droplets in various climatic ambient conditions**

Figure 11 demonstrates the influence of ambient conditions on droplet dispersion within the elevator for both the ventilation scenarios. For the quiescent scenario, larger droplets settle down in cold humid environments due to very low evaporation rates indicated by their relatively larger sizes by Fig. 11 (a), and eventually, a considerable fraction reaches the elevator floor within the specified elevator-travel duration of 10s. In contrast, findings for hot dry environments are obtained where, due to relatively high evaporation rates, the sizes continuously decrease significantly. As seen in the diameter distribution plot of Fig. 12, the droplets remain suspended in the domain due to their small masses. Furthermore, they become trapped in the flow field generated by both the cough and the constant inhalation and exhalation of the Covid-infected individual and spread across the elevator as shown in Fig. 12. On the other hand, in the cold-humid environment, in spite of being trapped in the flow field generated by human cough and the inhalation and exhalation, the droplets eventually settle down owing to negligible evaporation. Figure 12 also shows the mean diameter at various times, thus rendering a general overview of the total evaporation process. As can be observed from Fig. 12 for cold humid ambient, there is a slight increase in mean diameter initially, followed by a monotonic decrease. This can be explained by the phenomenon of supersaturation at high humidity, as reported by Chong et al[46].

In the fan ventilation scenario, circulation caused by the rotational effect increases the dispersion in droplets due to increased motion (produced by the rotating component of the fan), causing droplets to disperse to far-off locations and eventually get stuck at various elevator surfaces or fall below the inhalation breathing box height within a short span. The air-flow pattern brought about by the fan ventilation scenario at 1100 rpm outweighs major changes brought about by climatic influences on droplet dynamics. Hence climatic ambient conditions and their corresponding evaporation characteristics have less effect in dictating the droplet dispersion in the fan ventilation scenario. For all climatic conditions, a fan speed of 1100 rpm is maintained. Figure 12 demonstrates the diameter distributions of the suspended droplets for the fan ventilation scenario. It reconfirms the fact that a significantly less fraction of injected droplets remain suspended in the domain for the fan ventilation scenario. The diameter distributions demonstrate the effect of ambient conditions on droplet evaporation as indicated by the increase in droplet nuclei percentage with the change of ambient conditions from cold humid to hot dry. Although it has been concluded that climatic ambiance does not have a significant impact on the droplet dispersion in the fan ventilation scenario, Fig. 12 confirms that the fraction of suspended droplets in the hot dry ambience is greater than that of the cold-humid ambience owing to the difference in their evaporation characteristics as discussed above.

The size distribution plots for both the quiescent and fan ventilation scenarios (Fig. 12) depict the overall evaporation rate in a single plot for a particular scenario. For the scenarios involving dry climatic condition, the plots show that the majority of the droplets shifts towards the lower size range and close to 10s, most of the droplets are concentrated in the size range of 5-20 µm.

From the above discussions, we may conclude that in the quiescent scenario, the droplet size and evaporation rate which in turn is influenced by the ambient conditions, has a strong influence on the droplet trajectory whereas, in the fan scenario, the droplet size and evaporation rate (although following the same trend with the climatic condition as in the quiescent case) has less influence on the droplet trajectory and the droplet trajectory is controlled mostly by the airflow generated by the fan.

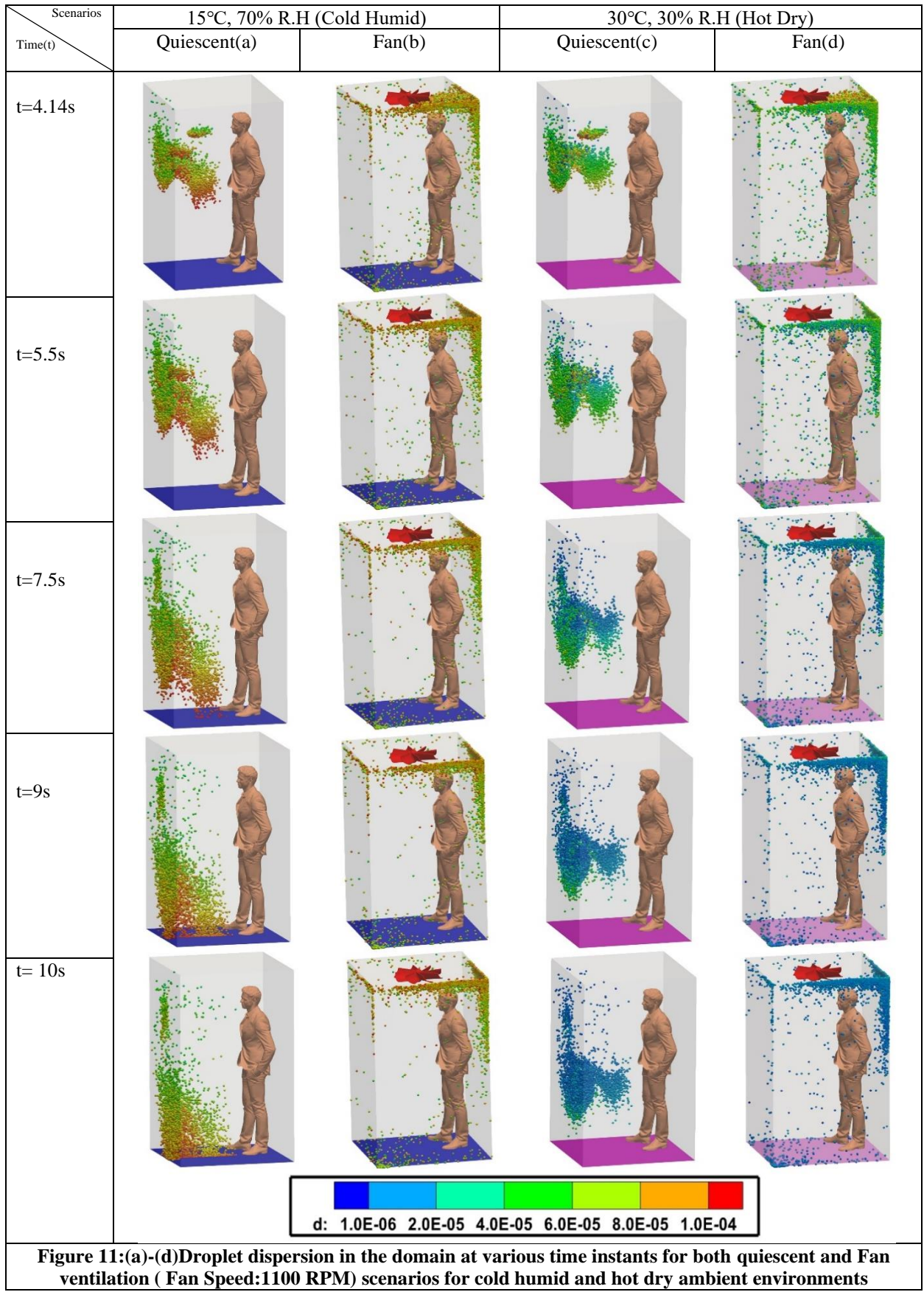

**Figure 11:(a)-(d)Droplet dispersion in the domain at various time instants for both quiescent and Fan ventilation ( Fan Speed:1100 RPM) scenarios for cold humid and hot dry ambient environments**

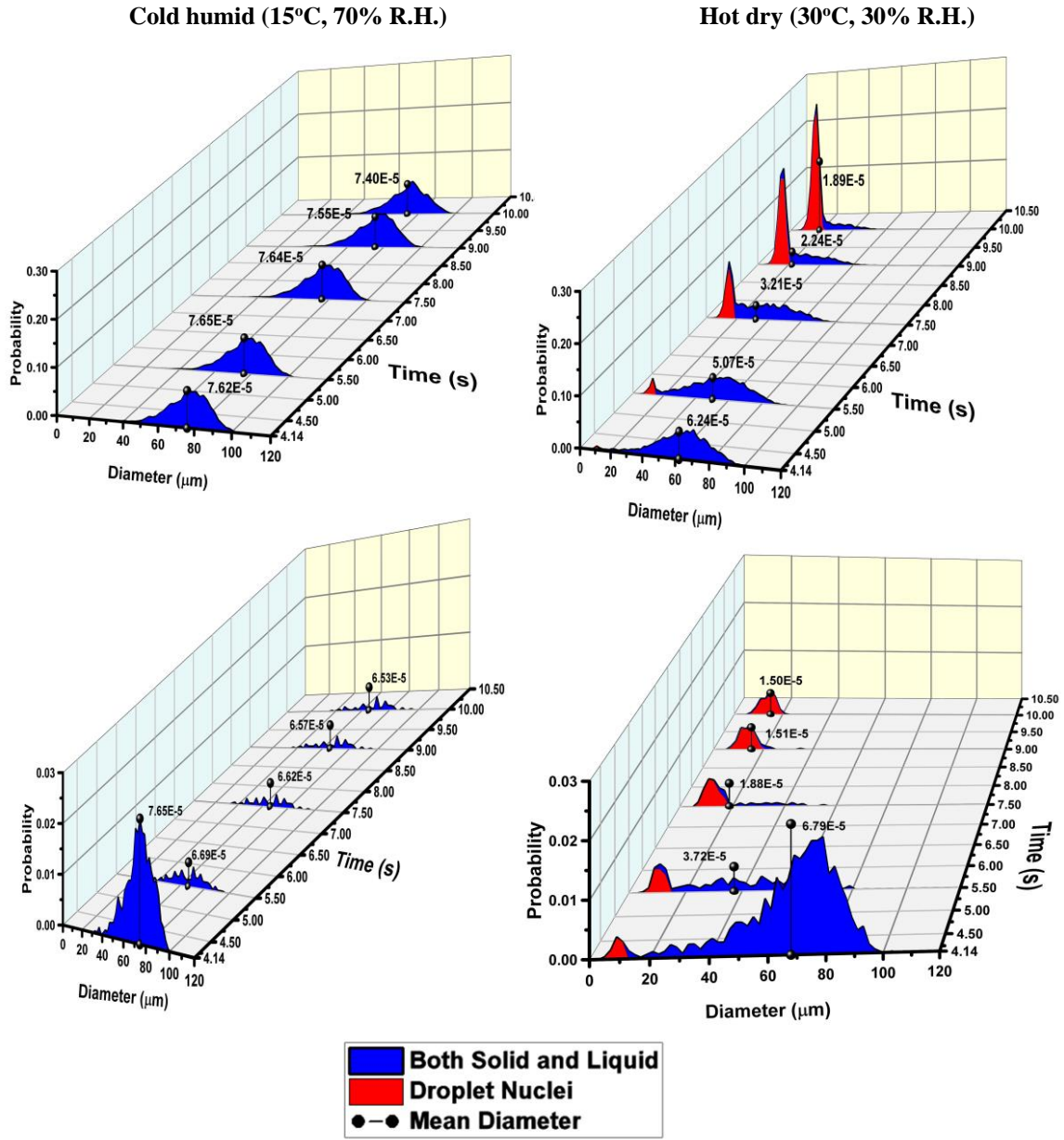

**Figure 12: Diameter size distribution in the domain at various time instants for Quiescent (above) and Fan (below) ventilation scenarios for cold humid (15°C, 70% R.H.) and hot dry (30°C, 30% R.H.) climatic ambiences respectively.**

**3.3.2. Epidemiological Implications in various climatic conditions**

Figures 13 (a-d) depict the spatial distribution of the risk of infection over the domain in various climatic conditions for the quiescent and fan ventilation scenarios respectively. In the case of a quiescent scenario, the risk is concentrated mainly in the region in front of the coughing person. This is because in this case, due to the absence of any external airflow, the droplets move forward by virtue of their injection velocity from the mouth, and descend slowly owing to gravity. Some droplets, entrained in the turbulence generated by the cough and the constant inhalation and exhalation of the passenger, spread across the elevator mainly towards the front right and left corners. Broadly speaking, in a quiescent scenario, the region in front of the mouth of the infected coughing person (i.e. the four breathing boxes in front) presents an immense risk of infection for all the climatic conditions, as shown by Fig. 13 (a-b). For the front left and right corners, the behavior is different for different climatic conditions. In the dry climatic condition case, due to more evaporation, the droplets which make their way to these spots remain suspended there for extended periods, whereas in the humid case, the droplets after reaching these spots, fall off the breathing box height quickly, thus explaining the difference.

For the Fan ventilation scenario, there is no such concentration of Risk in local regions as confirmed by Fig. 13 (c)-(d). This is because in the Fan ventilation scenario, due to the enhanced turbulence brought about by the rotational component of the fan, the droplets spread homogenously to various directions far away from each other and do not travel in a cluster (like they do in Quiescent scenarios). Furthermore, since a majority of droplets get deposited on the elevator surface rather than remaining suspended in the domain near the breathing box height as evidenced by Fig. 12, it ensures the fact that significantly low risk is maintained across the enclosed domain for the Fan ventilation scenarios.

The spatio-averaged risk of infection (i.e. the mean Risk of infection in the overall domain obtained by spatially averaging the individual Risk over all the 16 breathing boxes) and its comparison for the Quiescent and Fan ventilation scenarios is demonstrated in Fig. 14 for all climatic ambiences ( including hot humid(30°C, 70% R.H.) and hot dry(30°C, 50% R.H.)). In the Quiescent scenario, the cold humid ambient environment (15°C, 70% R.H.) has the lowest risk of infection followed by the hot humid ambient environment (30°C, 70% R.H.). The hot dry ambient environment (30°C, 30% R.H.) has the highest risk. As explained before, in cold humid ambient environments, the droplets owing to their large masses fall below the breathing boxes' height quickly and thus bringing down the risk significantly. The droplets in the hot-humid ambience also undergo negligible evaporation ( although more than cold-humid) and descend quickly ( quicker as compared to hot-dry and slower as compared to cold-humid) below the breathing boxes and thus bringing about a significant reduction in risk.

However, the droplets remain suspended for larger periods in the domain (also within the breathing boxes) owing to their smaller masses, in the hot dry ambient environment and produce significant risk. A similar trend of risk of infection according to the climatic conditions is observed for the fan ventilation scenario due to similar reasons (although the differences in risk due to change in ambient condition are not that pronounced as in the quiescent condition). However, in all the climatic conditions the risk corresponding to fan speed of 1100 rpm remains significantly low. Thus from Fig.14, we see that the Risk of Infection is highest in the hot dry climatic condition (30°C, 30% R.H.) and lowest in the cold humid climatic condition (15°C, 70% R.H.) and the Risk of Infection for the other climatic conditions lies between these two extremes for both the Quiescent as well as the Fan ventilation scenarios. This bolsters our previous claim and provides justification for presenting the results of only the two extreme climatic conditions.

Until now, we have conclusively established that changes in climatic environments have a major impact on droplet transmission routes and ultimately on the risk of infection in a quiescent scenario. It is worth emphasizing, however, that humidity is a more critical climatic aspect than temperature. Figure 14 depicts the spatially averaged risk of infection over the total exposure time as the climatic settings from cold humid (15°C, 70% R.H) to hot dry (30°C, 30% R.H) to hot humid (30°C, 70% R.H) for a quiescent scenario (shown in red color). Keeping the temperature (30°C) same, a change of 10.84% in average-risk factor is observed with the change in humidity (30% to 70% R.H.) whereas a change of 1.2% is observed with the change in temperature (30°C to 15°C) at same humidity level (70% R.H.).

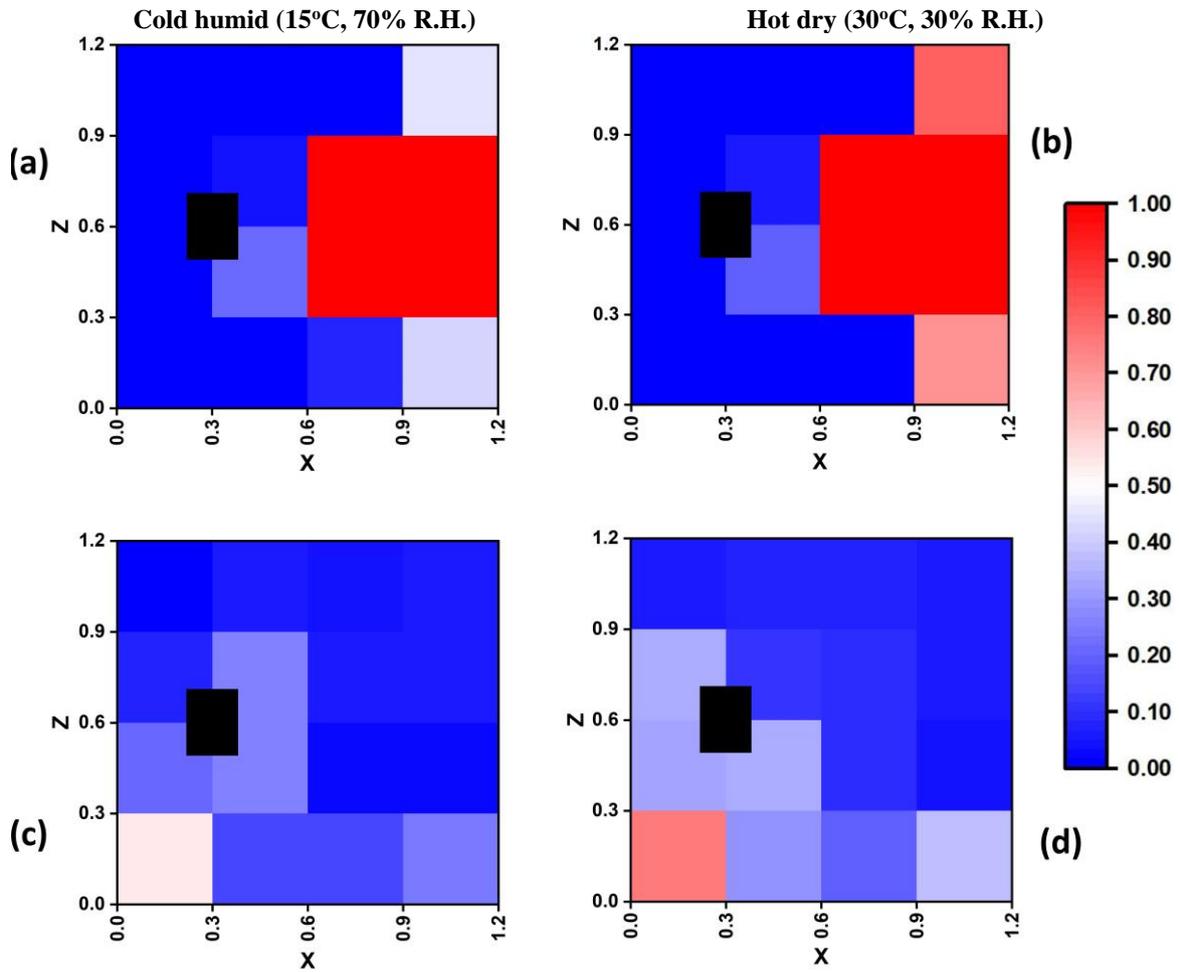

**Figure 13 (a)-(d):** Risk map showing the variation of risk of infection across the breathing boxes for Quiescent (above) and Fan (below) ventilation scenario throughout the domain over the total exposure time for cold humid (15°C, 70% R.H.) and hot dry (30°C, 30% R.H.) climatic ambiences respectively.

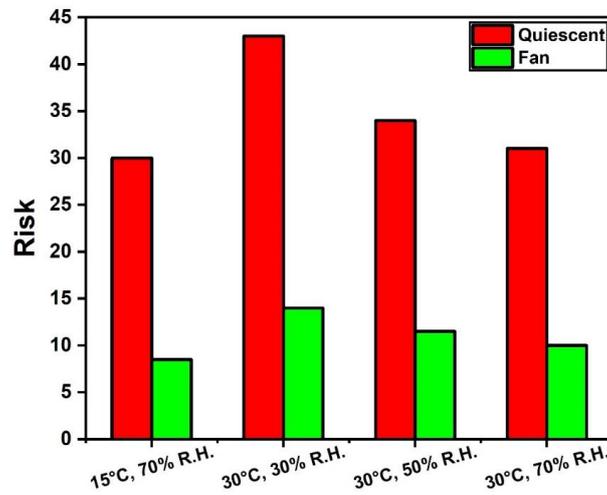

**Figure 14:** Showing the variation of spatio-averaged risk over the domain for various climatic ambiences for both Quiescent (in red) and Fan (in green) ventilation scenarios.

### 3.3.3. Epidemiological Implications of droplet size

We have studied epidemiological implications for various ventilation and climatic conditions and have quantified the spatial variation of the infection risk across the 16 breathing boxes, in various scenarios. The dose-response model quantifies the risk of infection based on the quantity of virion inhaled by a susceptible person. Up till now, all the studies have been based on the dose-response model. However, the severity and the ramifications of infection not only depend upon the number of virions inhaled but also significantly on the virion containing droplet size inhaled. Compared to a bigger droplet an evaporated smaller droplet containing the same virion quantity is more dangerous. The settling positions of the inhaled droplets vary within the respiratory tracts according to the size of the droplets. Smaller droplets, especially the virusols (droplet size < 20µm), may go into the alveolar and bronchial region but the larger ones may get arrested in the extrathoracic region[38]. If the droplets penetrate deep into the lower parts of the respiratory tract i.e. in the alveolar and bronchial regions, then the severity of the infection is significantly higher as compared to the droplets that get detained within the upper respiratory tract. Furthermore, infection in the lower part of the respiratory tract increases the risk of hospitalization and enhances the level of possibility of clinical intervention in a person significantly[40]. The possibility of lasting damage to the respiratory system also increases significantly[40]. Furthermore the size of virusols (droplet size < 20µm) cause them to linger in the domain for extended periods of time thus increasing the probability of inhaling them[52]. Additionally, the virion escape time is much less for small-sized droplets, the escape time is directly proportional to the droplet diameter, following which the speed of infection propagation is more in such types of droplets[39]. Hence, a detailed analysis pertaining to the Virusols (droplet size < 20µm) has been carried out.

The spatially averaged quantity of virusols (droplet size < 20µm) over the 16 breathing boxes as a percentage of suspended droplets (within the breathing boxes) for all time instants for all the climatic conditions for all the ventilation scenarios has been depicted in Fig.15. The cold humid condition (15°C, 70% R.H.) does not produce any significant quantity of virusols and hence not been represented in the plot of Fig.15. Since the evaporation rate of droplets in hot-humid ambience is higher than that of cold humid ambience a significant percentage of suspended droplets are converted to Virusols (droplet size < 20µm) in hot-humid ambient environments for the quiescent scenario as depicted in Fig.15. The hot and dry ambient in a quiescent scenario produces a significant percentage of virusols (droplet size < 20µm), owing to their significantly high evaporation rates, as can be inferred from Fig.15. This is due to the high evaporation rate in hot and dry ambient as evidenced by the diameter distribution plots in Fig. 12. However, the percentage of virusols (droplet size < 20µm) in the domain is low for all climatic ambient environments for fan ventilation because the percentage of suspended droplets in the breathing boxes within the appropriate height zone is inherently very low for the fan ventilation scenario in all climatic settings as can be seen from Fig.15. Thus we conclude a quiescent hot dry ambient condition leads to the formation of a significant quantity of suspended virusols (droplet size<20µm). Droplets in hot and dry ambient due to their relatively smaller sizes as compared to the other ambience will have a greater deposition probability in the alveolar and bronchial regions. The probable particle deposition count in the extrathoracic region $D_{c_{ET}}$ and in the bronchial and alveolar region $D_{c_{BA}}$ within any breathing, box is quantified through a formulation as per equation 3.

$$D_{c_{ET}} = \langle \Sigma P_{i_{ET}} N_{D_i} \rangle_{B_j} , D_{c_{BA}} = \langle \Sigma P_{i_{BA}} N_{D_i} \rangle_{B_j} \qquad (3)$$

; where $N_{D_i}$ is the total number of particles of diameter $D_i$ suspended within the breathing box $B_j$ and $P_{i_{ET}}$ and $P_{i_{BA}}$ are the deposition efficiencies[38] in the extrathoracic and bronchiolar and alveolar regions for that size of droplet respectively. Calculations using the above formulations are carried out for all the climatic conditions in a quiescent ventilation scenario. The respective temporally-averaged count of droplets within a breathing box that gets deposited within various parts of a respiratory tract, namely the extrathoracic and the alveolar and Bronchial regions, are depicted in Fig.16 for the Quiescent hot dry scenario. The cold humid, as well as the hot humid conditions, do not produce any droplets that might get deposited in the alveolar and bronchial regions. This is due to the fact that negligible evaporation occurs in the hot humid and cold-humid ambiences. On the other hand, extensive evaporation brought about by the hot dry ambience leads to the generation of a significant quantity of droplets that might get deposited in the alveolar and bronchial regions. The difference in evaporation characteristics and its impact on the evolution of the diameter of the suspended droplets has been depicted through the diameter ratio ( $\frac{D_i}{D_i^0}$; $D_i$: diameter of the ith droplet at a certain time instant, $D_i^0$: Initial diameter of the corresponding ith droplet) the plot of Fig.17. Thus it can be concluded that the hot dry scenario not only produces the highest risk of infection but also increases the probability of infection severity by producing a significant quantity of droplets that can deposit in the lower parts of the respiratory tract.

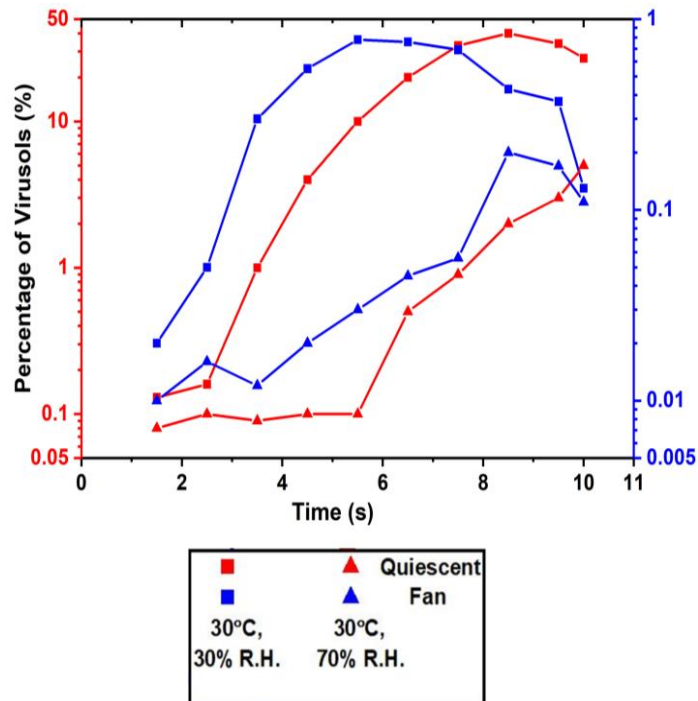

**Figure 15: Plot showing quantity of Virusols (droplet size < 20µm) as percentage of suspended droplets for Quiescent and Fan ventilation scenario in various ambient environments.**

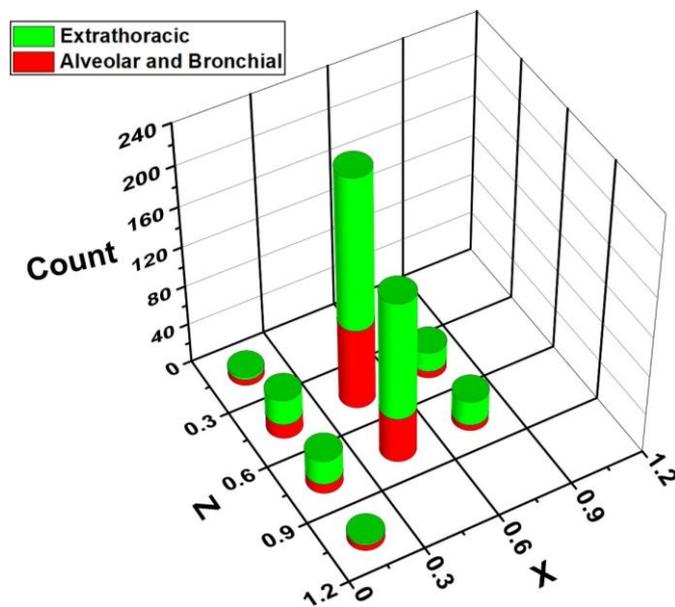

**Figure 16: Depicting the temporally-averaged count of droplets within each breathing box that might get deposited in various parts of the respiratory tract in a quiescent ventilation scenario for a hot dry ambience(30ºC,30% R.H.)**

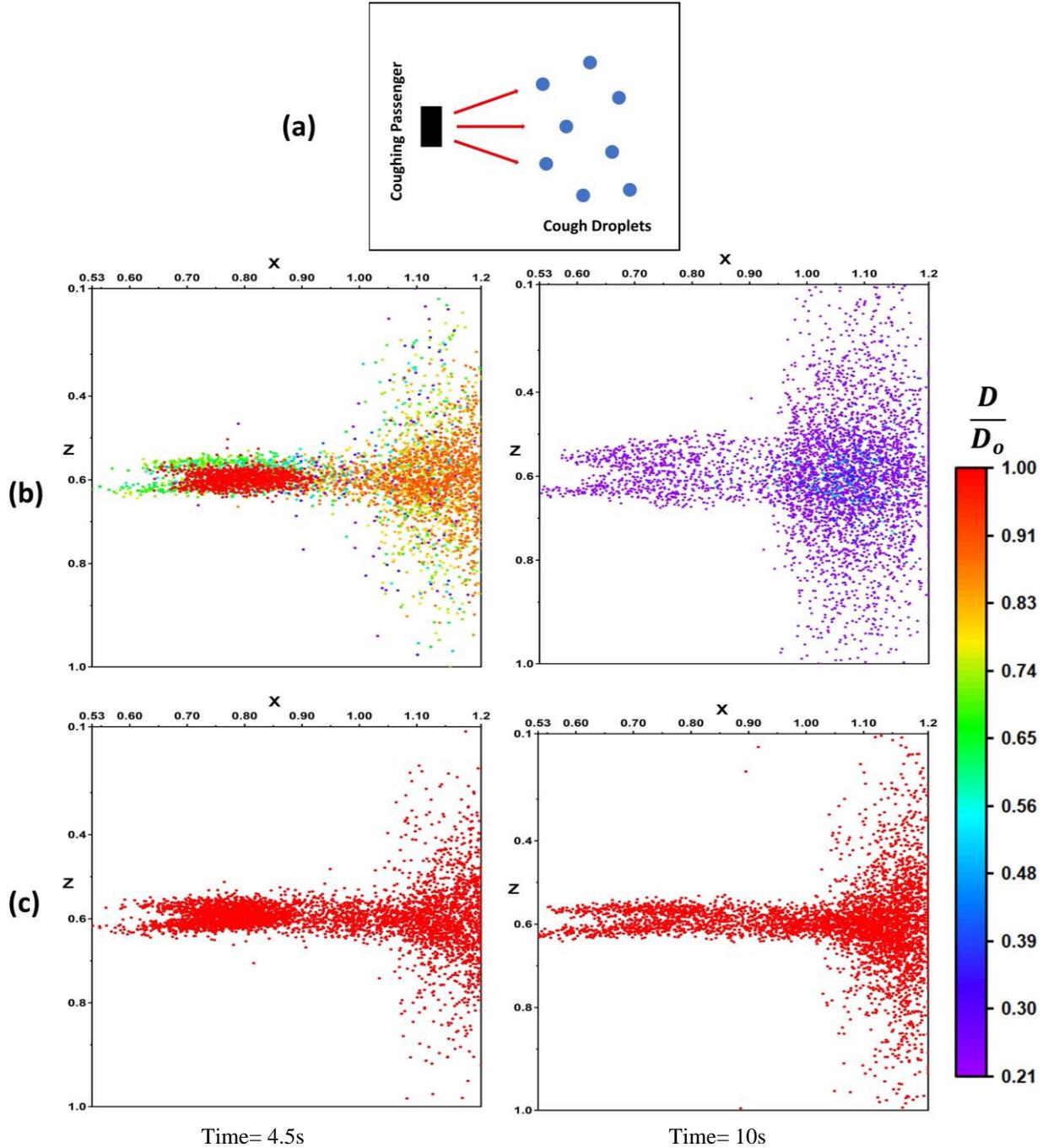

**Figure 17 (a): Schematic showing top-view of the coughing passenger emanating cough droplets into the domain.**

**(b-c): Plot showing Diameter ratio ($\frac{D}{D_0}$) of the cough droplets across the domain for quiescent scenario for hot dry condition (b) and cold humid condition (c) at various time instants.**

### 3.3.4. Infection Transmission through the Fomite route.

The discussion on infection transmission so far has been focused solely on the airborne transmission route. However, one more equally significant route of transmission, i.e. the fomite route of infection transmission exists for the coronavirus[34]. Whenever pathogenic particles are deposited on solid surfaces, the surfaces transform the particles into fomites. People who come into contact with these infectious surfaces may get the infection by transferring the pathogens through their hands. The droplet dispersion Fig.11 clearly demonstrates that a significant percentage of injected droplets get stuck on the elevator surfaces, mainly for the fan ventilation scenario. Hence these situations present a risk of infection through an indirect contact transmission route. Without bringing these transmission routes into account, the risk analysis would be incomplete. Thus a thorough investigation of indirect transmission is indispensable in understanding the risk of infection.

The risk of fomite infection in each of the four elevator walls (left, right, front, and back as shown in Fig. 18) and its variation with height has been evaluated using the Nicas and Best model[56]. Owing to the significantly less decay rate on the elevator surfaces (assumed to be steel) Nicas and Best[56] model has been implemented (equations 4-6) using relevant literature data[56-58] and the assumption that a person touches the elevator surface once during the elevator travel time.

$$E_m = f_m c_m A_s \overline{C_{\text{hand},t_o}} t_o \quad (4)$$

$$\overline{C_{\text{hand},t_o}} = \frac{f_b c_h C_s}{(\varphi + f_h c_h + f_m c_m) t_0} \left[ t_0 + \frac{\exp(-(\varphi + f_h c_h + f_m c_m) t_o) - 1}{\varphi + f_h c_h + f_m c_m} \right] \quad (5)$$

$$R = 1 - \exp(-\sigma E_m) \quad (6)$$

$E_m$ being the dose of pathogen delivered to the mucous membrane, $c_h$ is the pathogen transfer efficiency from the surface to the hand after a contact, $c_m$ is the pathogen transfer efficiency from the hand to the mucous membrane after a contact, $f_h$ is the frequency of hand-to-contaminated surface contact, $f_m$ is the frequency of hand-to-mucous membrane contact, $A_S$ is the average contaminated surface area touched per hand contact, $t_o$ is the concerned time interval, $C_s$ is the pathogen load per area of the contaminated surface, and $\varphi$ is the decay rate of the pathogen on hand.

The fan ventilation scenario produces a significant fomite infection risk for all the ambiences, however, the change in ambient conditions does not bring about any change in the fomite infection risk. The air-flow pattern brought about by the Fan ventilation scenario at an RPM of 1100 dominates over all major changes brought about by climatic influences on droplet dynamics and hence there is no change in droplet dispersion in the fan ventilation scenario and subsequently on fomite risk with the change in climatic conditions, whereas on the other hand, the quiescent scenario in all ambiences produces negligible fomite infection risk as enumerated in Fig.18. The walls are classified as left, right, front, and back with respect to the passenger inside the elevator who is facing the elevator door. The back wall (BW) and left wall (LW) pose the highest fomite risk of infection whereas the front wall (FW) and right wall (RW) present negligible fomite risk thus warning the passengers to avoid leaning on the back and right walls. The asymmetry between the left and right walls is mainly attributed to the flow circulation brought about by the anticlockwise rotation of the fan (as seen by the passenger from within the elevator). Hence, we conclude that although the fan ventilation scenario with 1100 RPM maintains significantly low risk through airborne transmission routes, it presents a substantial fomite transmission risk. Thus for the passengers, it is advised to take utmost precaution in abstaining from touching the walls as much as possible and to follow proper sanitization protocols after leaving the elevator.

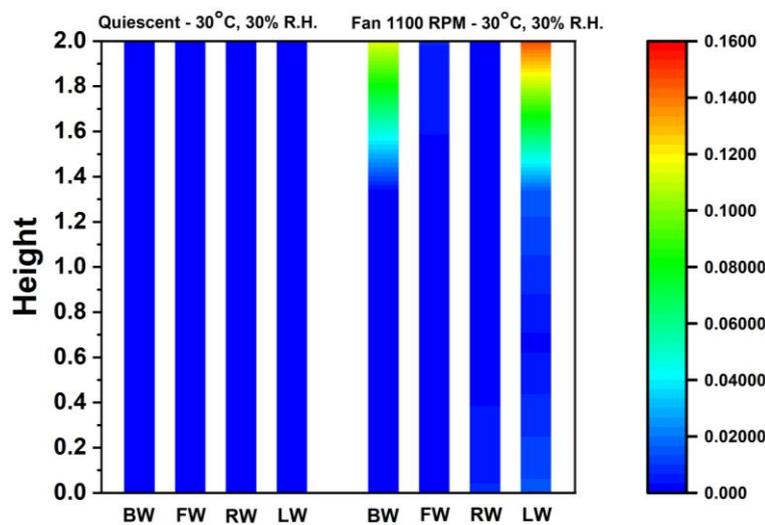

**Figure 18 : Contour plot showing the fomite risk variation with height on the elevator walls(BW: Back Wall, FW: Front Wall, LW: Left Wall, RW: Right Wall) for both quiescent and Fan ventilation scenarios.**

## 4. CONCLUSIONS

The dispersion evaporation characteristics and epidemiological implications of injected droplets in a small elevator typically used in small buildings or enterprises have been studied. The effect of ventilation scenarios (Quiescent and Fan) as well as climatic conditions (Cold Humid, Hot Dry, Hot-Humid) on droplet dispersion and evaporation has been investigated. The proper Fan speed that ensures minimum risk in all climatic conditions has been determined.

A dose-response model has been implemented for investigating the spatial variation of airborne risk of infection in the domain based on the pathogen count within the breathing zone of a human (16 boxes, each of size 0.3mX0.3mX0.4m, at a height of 1.45m). The results suggest that a quiescent state in the elevator carries a very high risk because a substantial fraction of droplets stay suspended in the domain (especially in the vicinity of the human head), particularly in the hot dry condition (30°C, 50% R.H and 30°C, 30% R.H). The airborne risk of infection over the total exposure time was 35.68% and 36% for 30°C, 50% R.H and 30°C, 30°C% R.H ambient environments respectively. In contrast, the cold humid condition (15°C, 70% R.H) have a lower risk factor as the droplets have larger masses throughout the travel time owing to negligible evaporation and quickly settle down below the height of the human breathing zone. (risk of infection, 32%). The risk of infection in a hot humid ambient environment (30°C, 70% R.H) lies in between the cold humid, and hot dry ambient environments (risk of infection, 33%). The evaporation rate although lower as compared to hot dry ambient is higher as compared to the cold humid environment. Hence as compared to a hot dry ambient environment, droplets descend quickly below the breathing zone height. However, since the evaporation rate is higher as compared to a cold humid ambient environment, more droplets remain suspended in the domain. It was further established that although both temperature and humidity have a significant effect on the risk of infection, humidity has a more pronounced effect on the risk factor as compared to temperature.

The introduction of forced convection in the form of a fan, at proper fan speed, alleviates the condition for all climatic conditions. The Fan RPM of 1100 turns out to be successful in significantly lowering the airborne risk as compared to the quiescent scenario at all instants in the domain for all ambient conditions, by 8% to 14% for cold humid, and hot dry ambient environments respectively, and increasing the Fan speed beyond 1100 RPM does not yield significant risk reduction for any of the above climatic conditions. It was observed that at this fan speed of 1100 RPM, climatic influences cannot exert major changes in droplet dynamics and the risk arising out of this. Furthermore, it was noticed that although the Fan ventilation ensures a low airborne risk of infection in the domain, it increases the Fomite risk of infection significantly in all climatic conditions (Maximum risk of 16% in all climatic conditions) as compared to the quiescent scenario which has negligible fomite risk in all climatic conditions. Hence, although Fan ventilation needs to be maintained to minimize the more dangerous airborne risk of infection, precautionary measures involving avoidance of touching surfaces or an after-travel sanitization must be maintained to nullify the fomite risk.

Another important aspect that was investigated, was the development of small-sized droplets in the diameter range of less than 20 μm. These virusols (droplet size<20μm), small-sized droplets with high viral loads have the highest penetration power in the alveolar and bronchial region of the respiratory tract and hence their transmission characteristics within the breathing boxes need to be investigated. In the quiescent ventilation scenario, the hot dry ambient environments due to their very high evaporation rate produce a significant quantity of Virusols (droplet size < 20μm) (24% of suspended droplets in the breathing zone at the end of 10s are virusols (droplet size<20μm)). The cold humid and hot humid ambient environments due to their significantly low evaporation rates produce negligible quantities of virusols (droplet size<20μm). The fan ventilation scenario due to the inherent low-risk factor produces a significantly low quantity of virusols (droplet size<20μm) in the breathing zone. The hot dry quiescent situation produces not only a situation with a very high likelihood of infection but also develops droplets of size ranges that have the maximum potential of penetrating into the bronchial and alveolar region. The spatial variation of the probable risk of infection in the extrathoracic as well as in the alveolar and bronchial regions of the respiratory tract has been thoroughly studied for all climatic ambiences in a quiescent ventilation scenario. The probable risk of infection in the alveolar and bronchial region for the cold humid and hot humid ambience is negligible whereas this type of infection risk is significant for hot dry conditions. Furthermore, for hot dry ambience, it has been concluded that the probable risk of infection in the alveolar and bronchial regions increases as one moves further from the afflicted person. Thus, one can safely conclude hot dry climatic ambient environment in a quiescent scenario is the most dangerous climatic condition in terms of covid transmission.

ACKNOWLEDGMENT

We want to express our gratitude to the High Performance Computing Cluster at Jadavpur University's Technological Bhavan for helping us to complete the simulations in a timely manner.